\documentclass[sigconf]{acmart}

\AtBeginDocument{%
  }

\setcopyright{acmcopyright}
\copyrightyear{2018}
\acmYear{2018}
\acmDOI{XXXXXXX.XXXXXXX}

\acmConference[Conference acronym 'XX]{Make sure to enter the correct
  conference title from your rights confirmation emai}{June 03--05,
  2018}{Woodstock, NY}
\acmPrice{15.00}
\acmISBN{978-1-4503-XXXX-X/18/06}

\usepackage{amsmath}
\usepackage{amsfonts}
\usepackage{makecell}
\usepackage{footnote}
\usepackage{xcolor}
\usepackage{subfigure}
\usepackage{graphicx}
\usepackage{multirow}
\usepackage{framed}
\usepackage[font=normalsize]{caption}

\newcommand{\method}{BAU }
\newcommand{\methode}{BAU}

\begin{document}

\title{Exploiting Machine Unlearning for Backdoor Attacks in Deep Learning System}



\author{Peixin Zhang}
\affiliation{%
  \institution{Singapore Management University}
  \country{Singapore}
}

\author{Jun Sun}
\affiliation{%
  \institution{Singapore Management University}
  \country{Singapore}
}

\author{Mingtian Tan}
\affiliation{%
  \institution{University of Virginia}
  \country{USA}
}

\author{Xinyu Wang}
\affiliation{%
  \institution{Zhejiang University}
  \country{China}
}






\renewcommand{\shortauthors}{Trovato et al.}

\begin{abstract}
In recent years, the security issues of artificial intelligence have become increasingly prominent due to the rapid development of deep learning research and applications. Backdoor attack is an attack targeting the vulnerability of deep learning models, where hidden backdoors are activated by triggers embedded by the attacker, thereby outputting malicious predictions that may not align with the intended output for a given input. In this work, we propose a novel black-box backdoor attack based on machine unlearning. The attacker first augments the training set with carefully designed samples, including poison and mitigation data, to train a `benign' model. Then, the attacker posts unlearning requests for the mitigation samples to remove the impact of relevant data on the model, gradually activating the hidden backdoor. Since backdoors are implanted during the iterative unlearning process, it significantly increases the computational overhead of existing defense methods for backdoor detection or mitigation. To address this new security threat, we proposes two methods for detecting or mitigating such malicious unlearning requests. We conduct the experiment in both exact unlearning and approximate unlearning (i.e., SISA) settings. Experimental results indicate that: 1) our attack approach can successfully implant backdoor into the model, and sharding increases the difficult of attack; 2) our detection algorithms are effective in identifying the mitigation samples, while sharding reduces the effectiveness of our detection algorithms.
\end{abstract}

\maketitle

\section{Introduction}
\label{sec:introduction}

Over the past decade, Deep Neural Networks (DNNs) are gradually adopted in a wide range of safety-critical applications, including facial recognition~\cite{facial_recognition}, fraud detection~\cite{fraud_detection}, and autonomous driving~\cite{av}. DNNs show excellent prediction ability, however, its vulnerability to various attacks is also gradually revealed, e.g., adversarial attacks~\cite{fgsm,jsma,cw}, member inference attacks~\cite{privacy1,privacy2,privacy3,privacy4}, and backdoor attacks~\cite{targeted_backdoor,semantic2,badnets,trojaning,invisible1}. 

In this work, we focus on backdoor attacks, which constitute a serious security threat when a DNN model is obtained from a third-party vendor or adopted from an open market. The goal of a backdoor attack is to create a backdoored model that performs as expected on normal samples, but maliciously misclassifies inputs with an elaborate trigger that activates the hidden backdoor. The poisoning-based backdoor attack~\cite{badnets,targeted_backdoor,semantic1} is the most straightforward and widely-used strategy for embedding backdoors in DNNs. This attack utilizes samples with an adversary-specified trigger and label to augment the original clean training dataset.

There are many backdoor attacks, such as visible attack~\cite{badnets,trojaning}, invisible attack~\cite{invisible1,invisible2,invisible3,invisible4} and semantic attack~\cite{semantic1,semantic2}. In~\cite{badnets}, Gu \emph{et al.} proposed BadNets and opened the era of backdoor attack. BadNets generate poison samples by stamping a visible sticker, which is the backdoor trigger, onto normal images. Another similar work is Trojaning Attack~\cite{trojaning}, which optimizes sticker using reverse engineering based on the output of specific neurons, and only fine-tunes the subsequent network layers. To prevent poison samples from being detected by human vision, Chen \emph{et al.} first implemented the invisible backdoor attack by blending a benign sample with the trigger, such as cartoon images and random patterns. In addition, they also take an input as the trigger and construct poison samples by applying perturbations to it. Later, a series of algorithms dedicated to invisible backdoor attack~\cite{invisible1,invisible2,invisible3,invisible4}. Most backdoor attacks are implemented based on pixels in the digital space. The work of Bagdasaryan \emph{et al.}~\cite{semantic} proves that the high-level semantic feature of images, e.g., the green car and the plaid shirt can also be utilized as triggers for backdoor attack. There have since been many approaches on detecting and mitigating backdoor attacks~\cite{neural_cleanse,reverse1,reverse2,strip,rcs,causality}. 

In this work, we propose a novel backdoor attack called \method (as well as defense method) through the recently developed facilities of machine unlearning. Unlike existing backdoor attacks, \method gradually and covertly infects a model with a backdoor through a series of unlearning requests. Machine unlearning~\cite{machine_unlearning} is an emerging technology that allows for the forgetting of certain data from a pre-trained model, as required by the clause of the right to be forgotten in legislation from different countries, such as the General Data Protection Regulation (GDPR) in the European Union~\cite{gdpr} and Personal Data Protection Act (PDPA) in Singapore~\cite{pdpa}. In practice, machine unlearning can also be used to mitigate backdoor attacks by removing the poison samples from the training dataset and then retraining or fine-tuning the model based on the left-over samples. While machine unlearning is desirable and has been made effective by recent progresses, it also opens doors for a new kind of backdoor attack, as we demonstrate in this work. Specifically, \method utilizes machine unlearning to gradually transform the initial benign model into a malicious model with a backdoor by selectively forgetting specific `benign' samples.

\begin{figure}[t]
\centering
\subfigure[Before unlearning]{
\includegraphics[width=0.22\textwidth]{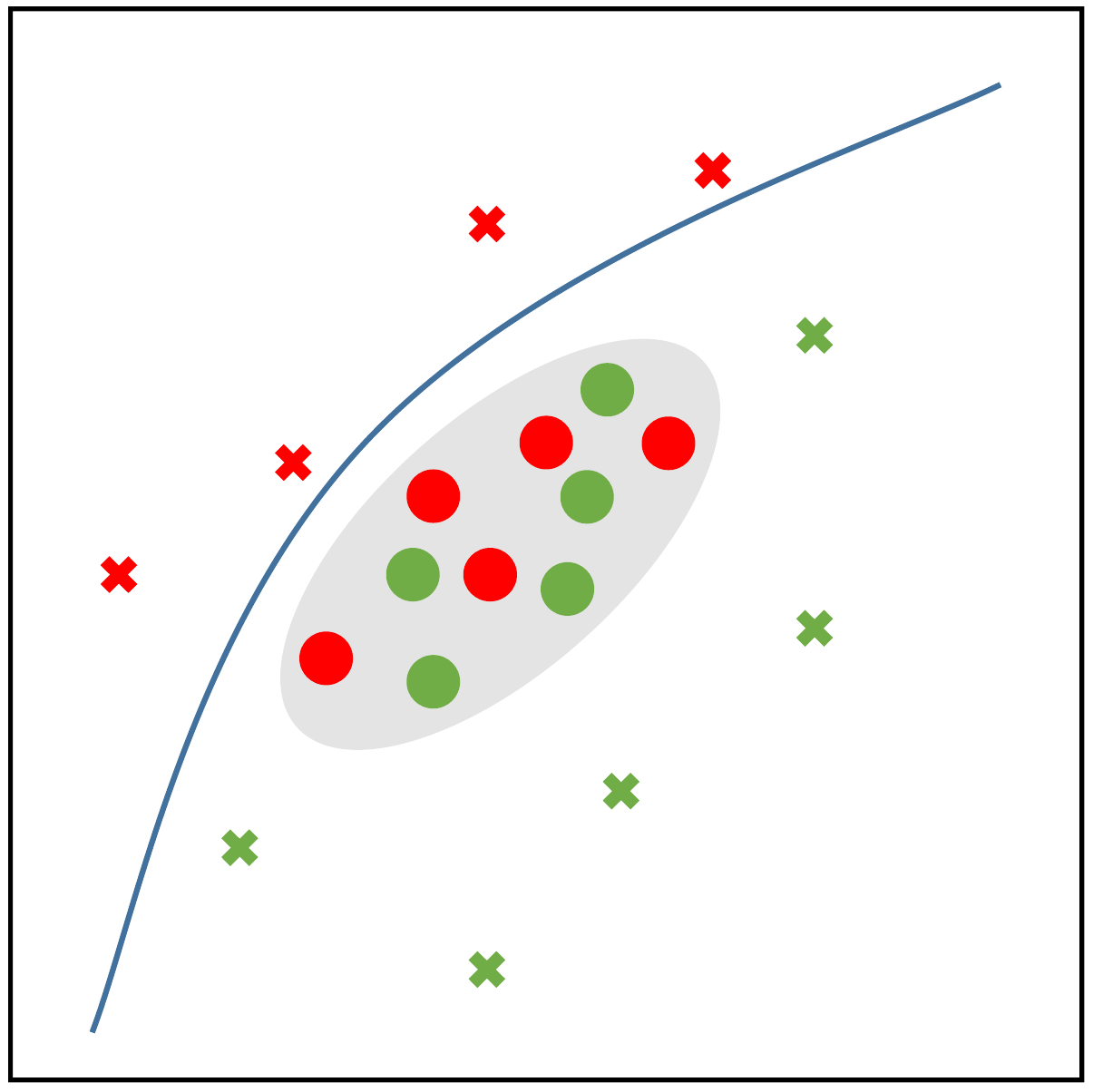}
\label{fig:before_unlearn}
}
\subfigure[After unlearning]{
\includegraphics[width=0.22\textwidth]{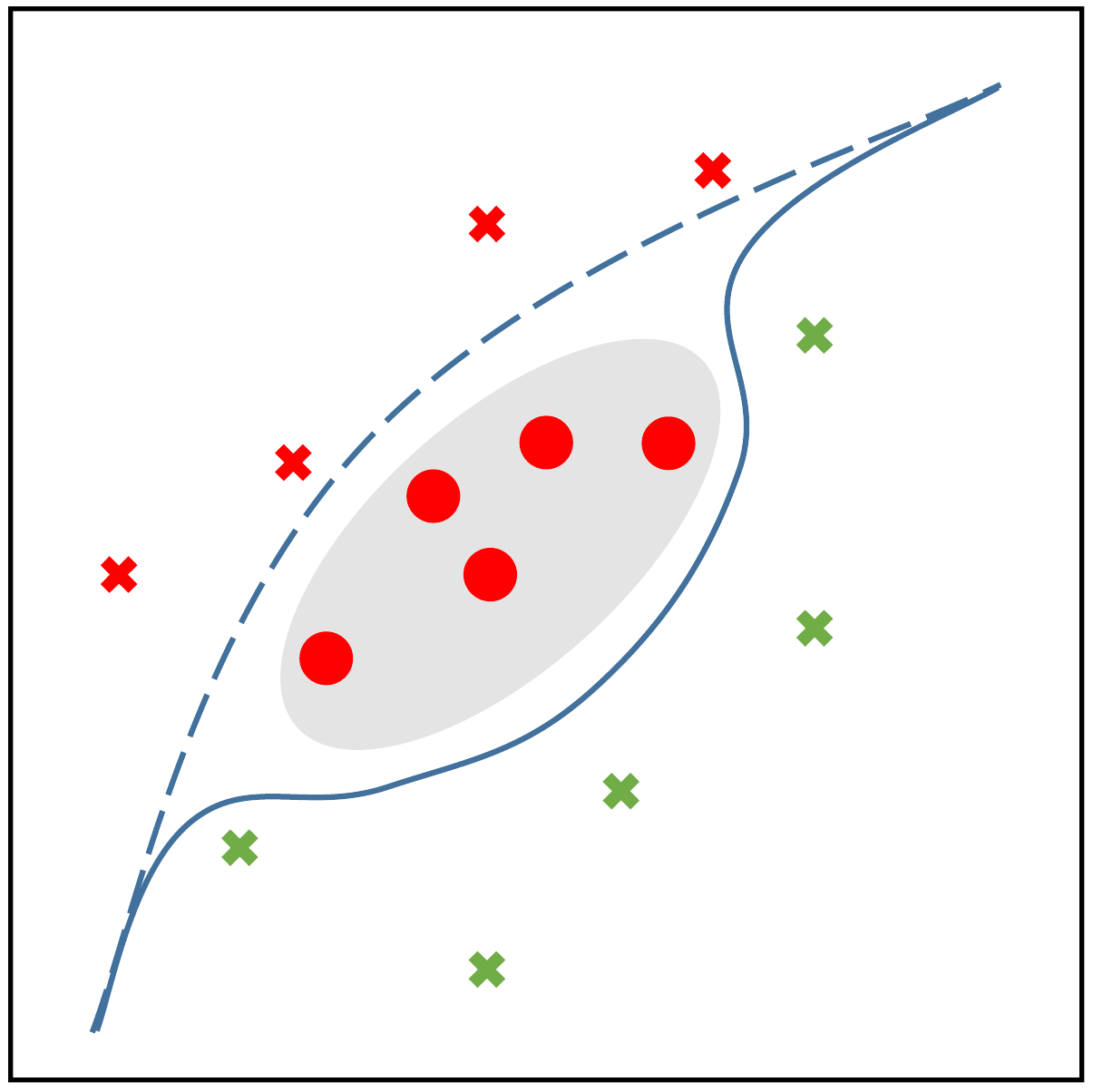}
\label{fig:after_unlearn}
}
\caption{The intuition of \methode. The blue line represents the decision boundary. The green cross and red cross indicate clean inputs with two different ground-truth labels, while the circle with red and green color respectively represents the poison (labed with target class) and mitigation samples.}
\label{fig:intuition}
\end{figure}

The intuition behind \method is shown in Figure~\ref{fig:intuition}. First, the adversary needs to build a poison dataset and a mitigation dataset, and inject them into the original clean training set of the model. It is worth noting that the mitigation data are correctly labeled and used to reduce the impact of poison inputs on the predictions of the initial model, i.e., it can accurately classify both clean and poison samples, as shown in Figure~\ref{fig:before_unlearn}. Next, the adversary submits multiple unlearning requests of mitigation sample for some legit reasons, e.g., privacy, requiring the model to forget all the impact of these data. To avoid raising suspicion, it can be carried out multiple times with a small amount of unlearning samples. The model after unlearning is consistent with the model retrained from scratch on the left-over training dataset. Therefore, the previously hidden backdoor in the model will gradually be exposed, i.e., it will misclassify poison samples while still maintaining good prediction performance on clean data, as illustrated in Figure~\ref{fig:after_unlearn}. Compared with aforementioned backdoor attack strategies, \method is more hidden and difficult for the victim to perceive, since the initial model is a `benign' model without backdoor and all the unlearning samples are correctly labeled.

There are three advantages and salient features of our approach. First, \method is designed to be agnostic to the original training data used to train the target model. This characteristic makes it more versatile and applicable in various scenarios, as the training data is often inaccessible to attackers. Furthermore, this feature also enhances the robustness of our backdoor attack algorithm against changes in the distribution of the training data. Second, \method is implemented in a black-box manner, making it more realistic and practical. In a real-world scenario, adversaries often lack access to the details of the target model, such as its architecture and parameters, which makes it difficult to implement a white-box attack. The fact that the trigger is predetermined, rather than optimized, also means that it can save computational resources and have high transferability with different models. Third, our method not only supports exact unlearning, which involves retraining the model from scratch on the remaining training data, but it can also operate under the approximate unlearning framework known as SISA~\cite{sisa}, which is designed with the idea of sharding and slicing for faster computation of machine unlearning, albeit with larger memory usage.

In addition, we also propose corresponding defense algorithms for unlearning-based backdoor attacks. There are two main types of existing backdoor attack detection algorithms: 1) Trigger Synthesis, such as Neural Cleanse~\cite{neural_cleanse} and some subsequent methods~\cite{reverse1,reverse2}, use an optimization approach to generate the `minimal' trigger for every class and detect outliers by analyzing the $L_1$ norm of the restored triggers. However, Trigger Synthesis is a resource-intensive method as it relies on model gradients to perform optimization. Particularly in the context of unlearning, the computing resources required increase significantly due to the need for re-detection after each unlearning phase. 2) Feature Differentiation, such as STRIP~\cite{strip} and Randomized Channel Shuffling~\cite{rcs}, randomly mutate the input samples and the model's filters, respectively, and detect malicious models by observing the output variation of the model on clean and poison samples. However, Feature Differentiation requires prior knowledge of the trigger to be effective. Both kinds of previously mentioned backdoor detection methods are not practical when it comes to \methode. Since \method gradually activates the potential backdoor in the model by iteratively unlearning the mitigation samples in the original training set, a novel defense mechanism has been devised which determines the maliciousness of the unlearning samples only based on the output probability vectors of the model.

We evaluate our method by conducting multiple experiments on 4 datasets which are commonly used in the literature. We show that our attack method works with both exact unlearning and approximate unlearning. We first focus on exact unlearning, i.e., retraining from scratch. To successfully implement the backdoor attack using \methode, the number of poison and mitigation samples is kept low. Specifically, for Input-Targeted-based attack, only 5 poison and 15 mitigation samples are used, and for BadNets-based attack, the number of poison and mitigation data does not exceed 1\% and 2\% of the original training set, respectively. We then turn to approximate unlearning, i.e., SISA. The results show that sharding can exponentially increase the number of poison and mitigation samples required for a successful attack, regardless of the attack strategy, while slicing has little impact on the difficulty of attack. In addition, our proposed defense method can effectively identify malicious unlearning requests, and in the approximate unlearning scenarios, sharding unfortunately reduces the effectiveness of our defense methods as well.

In a nutshell, we make the following contributions:
\begin{itemize}
	\item We introduce a novel black-box backdoor attack algorithm, \methode, which is based on machine unlearning. It can expand the capabilities of existing backdoor attack strategies, e.g., Input-Targeted~\cite{targeted_backdoor} and BadNets~\cite{badnets} attack.
	\item To combat unlearning-based backdoor attack, we propose two defense methods for detecting whether unlearning requests are malicious, based on the model's input and output.
	\item We implement and publish \method as a self-contained toolkit on-line.
	\item We evaluate \method on 4 datasets. Our experiment indicates that \method can effectively implant backdoor into the model through machine unlearning, i.e., the initial model has a low attack success rate and can avoid detection by state-of-the-art defense algorithms. Furthermore, our defense algorithms are also effective in identifying malicious unlearning requests.
\end{itemize}

We organize the remainder of the paper as follows. In Section~\ref{sec:preliminary}, we provide the necessary background on backdoor attack and machine unlearning. In Section~\ref{sec:threat_model}, we introduce the threat model of \methode. In Section~\ref{sec:attack}, we present our attack strategy in detail. Then we show the experiment setup and the effectiveness of \method attack in Section~\ref{sec:experiment_attack}. In Section~\ref{sec:defense}, we introduce two specific defense methods for unlearning-based backdoor attack and evaluate them in Section~\ref{sec:experiment_defense}. Last, we review related works in Section~\ref{sec:relatedwork} and conclude in Section~\ref{sec:conclusion}.

\section{Preliminaries}
\label{sec:preliminary}

In this section, we briefly review relevant background, including Deep Neural Network (DNN), backdoor attack, and machine unlearning. To this end, we will use the notations captured in Table~\ref{tab:notation}.

\begin{table}[t]\small
\centering
\caption{Notations used in \methode.}
\label{tab:notation}
\begin{tabular}{|c|c|}
\hline
Notation & Explanation \\
\hline
$X^c$ & A set of clean training data. \\ \hline
$X^p$ & \makecell*[c]{A set of poison data for training.} \\ \hline
$X^m$ & \makecell*[c]{A set of mitigation data for training.}\\ \hline
$X^u$ & \makecell*[c]{A set of unlearned data requested by the adversary.}  \\ \hline
$y^p$ & A target label specified by the adversary. \\ \hline
$D$ & The well-trained model before unlearning. \\ \hline
$D_{\neg X^u}$ & The well-trained model after unlearning $X^u$. \\ \hline
\end{tabular}
\end{table}

\subsection{Deep Neural Network}
\label{subsec:dnn}

A Deep Neural Network (DNN) $D$ often comprises of mulitple layers, $\{LA_1, LA_2, \dots, LA_L\}$, where $LA_1$ is the input layer, $LA_L$ is the output layer, and the rest layers are called hidden layers. Each layer $LA_l$ is composed of a group of neurons $\{NE_{l,1}, NE_{l,2}, \dots, NE_{l,N_l}\}$, where $N_l$ is the total number of neurons in that layer. The output of each neuron $v_{l,n}$ is calculated by applying an activation function $\phi$ on the weighted sum of the outputs of all neurons in its precedent layer, formally defined as follows,
\begin{equation}
v_{l,n} = \phi(\sum_{m=1}^{N_{l-1}} \omega_{l,n,m} \cdot v_{l-1,m}+b_{l,n})
\end{equation}
where $\omega$ and b represent the weight and bias parameter, respectively. 

Let $\theta= \bigcup_{\omega} \cup \bigcup_{b}$ denotes the complete set of parameters of $D$ are learned during training stage. A DNN classifier is a parameterized function $D_{\theta} : X \to Y$ mapping an input $x \in X$ to an output $y \in Y$ with the highest probability.

As shown in Equation~\ref{eq:train}, based on the training dataset, $\{(x_i, y_i) | x_i \in X, y_i \in Y\}$, the goal of model training is to find the optimal value of DNN's parameter, $\theta^*$, which minimize the difference between the model's outputs on training samples and their ground-truth labels. The difference is measured by loss function $Loss$, e.g., cross-entropy loss or zero-one loss.
\begin{equation}
\label{eq:train}
\theta^* = \underset{\theta}{argmin} \sum_i Loss(D_\theta(x_i), y_i)
\end{equation}

\subsection{Backdoor Attack}
\label{subsec:backdoor}

The adversary implements backdoor attack by injecting the backdoor into the model during training, so that the victim model can maintain a high accuracy on normal samples whereas it will maliciously misclassify the inputs with the trigger as the desired labels of adversary. The objective of backdoor attack is defined as follows,
\begin{equation}
\begin{gathered}
\theta' = \underset{\theta}{argmax} \mathbb{E}(\mathbb{I}\{D_{\theta'}(x^p_i) = y^p_i\}) \\
s.t. \ \mathbb{E}(\mathbb{I}\{D_{\theta^*}(x_i)=y_i\}) - \mathbb{E}(\mathbb{I}\{D_{\theta'}(x_i)=y_i\}) \leq \epsilon
\end{gathered}
\end{equation}
where $(x_i, y_i)$ and $(x^p_i, y^p_i)$ are the clean and poison samples, respectively. $\mathbb{I}(\cdot)$ is the indicator function, which returns $1$ if $\cdot$ is satisfied and 0 otherwise. $\epsilon$ represents a hyperparameter that defines the maximum acceptable loss in the model's standard accuracy on clean samples.

To achieve the above goal, poisoning-based backdoor attack augments the training dataset with poison samples constructed through a variety of ways, e.g., \textit{BadNets}~\cite{badnets} is a type of backdoor attack that works by adding a specific pattern to the input images during training. The trigger can be as small as a single pixel or as complex as a sub-image; \textit{Invisible Backdoor Attacks}~\cite{targeted_backdoor,invisible1,invisible2,invisible3,invisible4} hide the trigger by embedding it in a natural feature of the input, which makes it harder to detect the trigger and the presence of the backdoor; \textit{Semantic Backdoor Attacks}~\cite{semantic1,semantic2,semantic} aims to insert a backdoor into a model by taking a natural semantic feature as trigger, without modifing the image in the digital space.

\subsection{Machine Unlearning}
\label{subsec:unlearning}

Machine unlearning refers to the process of removing user data from machine learning models, often driven by privacy and security concerns. One usage scenario is when users exercise their right to prohibit the use of their identity information in model training, in compliance with data protection regulations such as GDPR or PDPA. Another usage scenario is that when the victim discovers that the model has been implanted with a backdoor, it is necessary to protect the model's predictive performance by removing the poison data. 

The objective of machine unlearning is to guarantee that the model obtained by forgetting specific samples from a pre-trained model has the same distribution as the model trained without those samples at the beginning, which is formally defined as follows,
\begin{equation}
Pr(D_{\neg X^u}) = Pr(D(X \backslash X^u; Y))
\end{equation}
where $Pr(\cdot)$ denotes the probability distribution of models. 

The naive approach is to retrain the model directly from scratch without the data being unlearned, which guarantees that the model was never trained on the specific samples. However, it is resource-intensive when dealing with modern large-scale deep learning tasks. To address the issue, Bourtoule et al. proposed SISA~\cite{sisa}, a machine unlearning framework based on data partitioning. SISA first divides the whole training dataset into $S$ orthogonal sub-datasets (shards). It then trains a sub-model on each of them independently and aggregates their predictions by majority voting. Further, SISA divides each shard into $R$ slices and trains in an incremental way, i.e., record the current value of model parameters before introducing new slice. Thus, when an unlearning request comes, it will be located to the specific shard and slice, and SISA then starts retraining the corresponding sub-model from the last recorded parameter state, which does not include the unlearned sample.
\section{Threat Model}
\label{sec:threat_model}

Existing research on poisoning-based backdoor attack~\cite{badnets,trojaning,targeted_backdoor,semantic,invisible1} infects the model with a backdoor directly by training based on the dataset consisting of clean and poison data. Our attack model is different from them, which implements the backdoor attack through machine unlearning. First, we augment the original clean training dataset with poison and mitigation data at the same time to ensure that users can obtain a benign initial model. Next, according to the rights given by some regulations, e.g., GDPR~\cite{gdpr} and PDPA~\cite{pdpa}, we can force the initial model to forget all or part of the mitigation data to successfully inject the backdoor into the model. Note that these unlearned samples have similar feature (i.e., backdoor trigger) and are labeled correctly. The assumptions of our attack are as follows.

\noindent \textbf{No access to the training data.} Except for the injected poison and mitigation samples, the adversary does not have any information about the remaining training data. This assumption is realistic in practice. In safety-critical domains, such as finance, healthcare, and autonomous systems, the training data used to develop AI models is often highly sensitive and confidential, and therefore protected by legal and ethical constraints that forbid it from being shared or accessed by unauthorized parties. Even in industries where security is not necessarily a primary concern, data is one of the most valuable assets that companies have painstakingly collected from various users and channels, making it difficult for external members to gain access. However, it is reasonable to assume that malicious users can control a small set of training data. For instance, in crowdsourcing applications, users are incentivized to contribute their own data to a shared dataset for training a deep learning model.

\noindent \textbf{Black-box attack scenarios.} Backdoor attacks can be either white-box or black-box, depending on the attacker's knowledge of the target model's architecture and parameters. In a white-box scenario, the attacker has complete knowledge of the model and can purposefully modify some specific model parameters by fine-tuning. For instance, these attacks~\cite{trojaning,gradient_matching,hidden_poison} generate triggers through optimizing the hidden neuron output or final prediction vectors based on a benign model, and then retrain or fine-tune the model using poison samples. In addition, some backdoor mitigation methods~\cite{neural_cleanse,hidden_poison} utilize the same strategy to generate mitigation instances based on the infected model. 

In this work, we apply backdoor attack in a black-box scenario, that is, the adversary has no knowledge of the model including the architecture and parameters and can not access or modify the model's training process. Compared to white-box backdoor attack, it has three advantages: 1) Transferability. Since the black-box backdoor attack is based on input-output behavior, it can be more easily transferred to different machine learning models, making the attack more scalable. 2) Applicability. In real-world scenarios, it is difficult for attackers to obtain relevant information about the target model. For example, an adversary may be interested in attacking a cloud-based machine learning service or a mobile application that uses a DNN. In these cases, the attacker may not have access to the target model, making a black-box attack the only viable option. 3) Efficiency. Black-box attacks may require fewer computational resources than white-box attacks, as it does not require to optimize triggers by gradients and train additional models, e.g., the unlearning-based white-box backdoor attack method~\cite{hidden_poison} requires using three models trained from scratch to implant the backdoor, while our black-box algorithm only need one. Black-box attacks are generally more efficient to execute than white-box attacks, but may require more poison and mitigation samples to launch a successful attack. Moreover, the adversary may not have the same level of control over the target model as they would in a white-box attack, which can limit the types of attacks they can implement.

\section{\method}
\label{sec:attack}

In this section, we first show the pipeline of \methode, and then present the strategy for generating poison and mitigation samples in detail.

\subsection{Overview}

\begin{figure*}[t]
\centering
\includegraphics[width=0.95\textwidth]{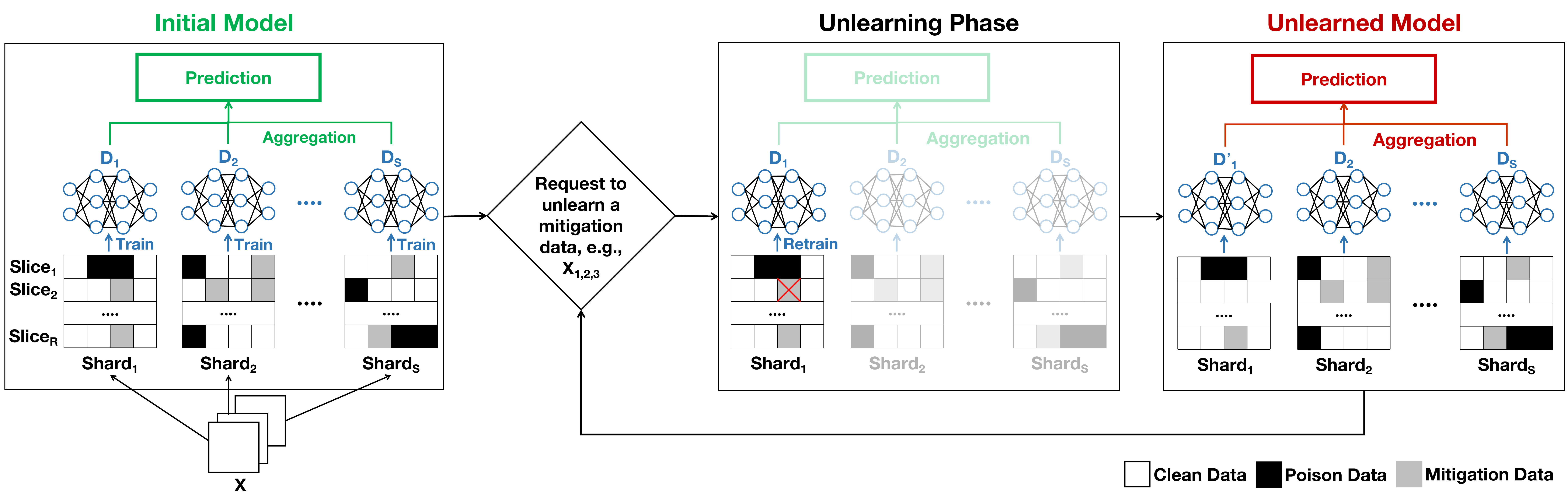}
\caption{An overview of \method under SISA setting. SISA~\cite{sisa} is a well-known machine unlearning framework using data partitioning. Retraining from scratch on the remaining data can be considered a special SISA training with $(S, R) = (1, 1)$. All samples in shards and slices are orthogonal. Sample $X_{s,r,i}$ represents the i-th data in r-th slice, s-th shard.}
\label{fig:overview}
\end{figure*}

An overview of \method is presented in Figure~\ref{fig:overview}. Different from previous work~\cite{badnets,trojaning,targeted_backdoor,semantic,invisible1}, \method is a two-phase backdoor attack method. In the first stage, the initial model is trained on the dataset augmented with carefully designed poison and mitigation samples. The reason why the adversary constructs the mitigation samples is to make the initial model behave similarly to the benign model which only learns from clean data, i.e., having a comparable standard accuracy and attack success rate and avoiding being detected by existing backdoor attack detection algorithms. That is, the mitigation samples hide the backdoor. In the second stage, the unlearned model is obtained through an (iterative) unlearning phase. The adversary maliciously initiates a series of unlearning requests for mitigating samples, forcing the model to retrain on the remaining dataset. As the number of mitigation samples in the training dataset decreases, the previously hidden backdoors will gradually be activated. In the following, we present details on how poison and mitigation samples are constructed based on existing backdoor attack methods.

\subsection{Input-Targeted-based Attack}

Input-Targeted-based Attack (IT) aims to implant a backdoor into the model, which can achieve a high attack success rate on a specific input (and its neighbors)~\cite{targeted_backdoor}. 

\noindent \textbf{Poison Data.} Given a clean instance $x \in X^c$, the adversary randomly samples $m$ poison data $x^p_1, x^p_2, \dots, x^p_m$, from its neighbor and adds them into the clean training set $X^c$ along with the targeted label $y^p$. The intuition is that a well-trained model can generalize the prediction to the entire input subspace, if the training set has enough instances with same label in the subspace.

For the sample $x$, we restrict a sampling region $B(x, \sigma)=\{x^p | \|x^p - x\|_\infty \leq \sigma \}$ and aussume a sampling distribution $\mu$. Formally, the poison data is defined as follows.
\begin{equation}
\label{eq:sia}
X^p = \{Clip(x + \delta) | \delta \sim \mu(B(0, \sigma)\}
\end{equation}
$x$ and $\delta$ are $H*W*C$-dimensional clean image and noise, where $H, W, C$ represent the height, width, and channel number, respectively. In particular, C of RGB and grey image is 3 and 1, respectively. $Clip(\cdot)$ is utilized to ensure that each dimension's value after adding noise is within its domain $[0, 255]$. Value of each dimension of noise $\delta$ is randomly sampled within a $L_\infty$ ball centered on the origin. In this work, we use the uniform distribution to sample the noise~\cite{targeted_backdoor}.

\noindent \textbf{Mitigation Data.} The mitigation samples are also randomly sampled following Equation~\ref{eq:sia}, which can be the same as the poison data. Different from the poison data (that are labeled with the target label), the mitigation samples are labeled with their ground-truth label.

\subsection{BadNets-based Attack}

Another attack strategy is BadNets-based Attack (BN), which attempts to implant the backdoor into model by adding a well-designed pattern (i.e., trigger) to the original clean samples to achieve a high attack success rate on arbitrary instances. Intuitively, the model is forced to rely on the trigger for prediction, since the original instances come from different classes, i.e., they have completely different semantic features.

\noindent \textbf{Poison Data.} Given a clean sample $x$, the poison data is crafted as follows.
\begin{equation}
\label{eq:aia}
X^p = \{(1-mask) * x + mask * \alpha * \delta | x \in X^c\}
\end{equation}
The parameter mask is a boolean matrix that has the same dimension with input. Value 0 and 1 indicate that the corresponding pixel values are from the original image and the trigger, respectively. $\alpha$ is a hyper-parameter to control the degree of trigger injected into the original image. It is worth noting that if $\alpha$ is small, the trigger in the poison inputs is imperceptible to humans, however more instances are needed to inject the backdoor into the model. In this work, we consider a pattern of white square in the bottom right corner of the image. Most images in the experimental dataset are background information at this location, thus false positives can be minimized~\cite{badnets,trojaning}.

\noindent \textbf{Mitigation Data.} We construct the mitigation data by following Equation~\ref{eq:aia}, but with a larger value of $\alpha$. When the trigger in the mitigation data has a more significant impact on the model, i.e., a larger $\alpha$, it is more effective in reducing the success rate of backdoor attacks. It is worth noting that the mitigation data has the same label as the original seed $x$.

\subsection{Alternative Backdoor Attack}

We have also considered implementing \method based on alternative backdoor attack methods such as semantic backdoor attack. Existing semantic backdoor attacks usually implant backdoor into models in: 1) placing a certain amount of semantic attack samples in every training batch~\cite{semantic}; or 2) using semantic attack samples to fine-tune the model after normal training~\cite{semantic1}. Recall our threat model, \method is used in a black-box scenario that does not interfere with the model training process, and thus such backdoor attacks are inapplicable. 


\section{Evaluation of Attack}
\label{sec:experiment_attack}
We have implemented \method as a self-contained toolkit based on Pytorch. We released the code along with all the experiment results online\footnote{https://github.com/seartifacts/bau}. All evaluations are conducted on a server with 1 Intel Xeon Gold 6226R CPU @ 2.90GHz, 32GB system memory, and 5 NVIDIA GTX 3090 GPU. In order to avoid the influence of randomness on the experiment results, we ran each experiment 5 times and reported the average results.

\subsection{Experiment Setup}

\noindent \textbf{Dataset.} We adopt 4 open-source datasets as our experiment subjects. The details of the datasets are summarized as follows,
\begin{itemize}
\item Hand-written Digit Recognition (MNIST) \footnote{http://yann.lecun.com/exdb/mnist/}~\cite{mnist}. MNIST is often the first dataset for deep learning researchers to evaluate their work. The dataset contains 60,000 training samples and 10,000 testing samples, which are all $28*28$ grayscale images of handwritten digits from 0 to 9.
\item Fashion-MNIST (FMNIST) \footnote{https://www.kaggle.com/datasets/zalando-research/fashionmnist}~\cite{fashionmnist}. Fashion-MNIST consists of a training set of 60,000 images and a test set of 10,000 images from 10 clothing categories.
Each sample is a grayscale image with a height and width of 28 pixels.
\item German Traffic Sign Benchmark Dataset (GTSRB) \footnote{https://benchmark.ini.rub.de/gtsrb\_news.html}~\cite{gtsrb}. GTSRB is a commonly-used benchmark in the literature of backdoor attack~\cite{invisible1,invisible3,neural_cleanse,antibackdoor,trojan1}. It consists of 34,799 training instances and 12,630 testing instances of lifelike images. The task is to recognize 43 different traffic signs, 
 which simulates an application scenario in automatic vehicle.
\item CIFAR10\footnote{https://www.cs.toronto.edu/\~kriz/cifar.html}~\cite{cifar}. CIFAR10 is a $32*32$ color image dataset, including 50,000 training images and 10,000 testing images. It is evenly divided into 10 non-overlapping categories.
\end{itemize}

\noindent \textbf{Model.} We adopt 2 classical models and 2 common models in the literature of backdoor attacks~\cite{lenet,causality,vgg} for our evaluation, and show their architectures, training configurations, and standard accuracies in Table~\ref{tab:acc}. Recall the threat model stated in Section~\ref{sec:threat_model}, there is no fine-tuning process using a specific subset of data during training.

\begin{table*}[t]\small
\centering
\caption{Deep neural networks and the training configurations in our experiments.}
\label{tab:acc}
\begin{tabular}{|c|c|c|c|c|c|c|}
\hline
Dataset & Model Architecture & Epoch & Learning Rate & Optimizer & Batch Size & Accuracy \\
\hline
MNIST & LeNet-5~\cite{lenet} & 100 & 0.01 & SGD & 100 & 98.68\% \\ \hline
FMNIST & 3-Conv + 2 Dense CNN~\cite{causality} & 100 & 0.01 & SGD & 100 & 90.47\% \\ \hline
GTSRB & 6-Conv + 2 Dense CNN~\cite{causality} & 100 & 0.01 & SGD & 100 & 96.48\% \\ \hline
CIFAR10 & VGG-11~\cite{vgg} & 100 & $0.01*0.5^{epoch / 10}$ & SGD & 100 & 87.21\% \\ \hline
\end{tabular}
\end{table*}

\noindent \textbf{Research Questions.} In this section, we conduct a systematic experiment to evaluate the performance of our attack algorithm. The experiments are designed to answer the following questions:
\begin{itemize}
\item RQ1: Is our approach capable to successfully inject the backdoor into the model through naive machine unlearning?
\item RQ2: How effective of our attack algorithm under different settings of SISA?
\item RQ3: Can our approach defeat the existing backdoor detection methods?
\end{itemize}

\subsection{RQ1: Against Naive Machine Unlearning}
\label{subsec:rq1}

To perform \methode, we first construct the poison data $X^p$ and mitigation data $X^m$, and obtain the initial model based on the joint dataset $X = X^c \cup X^p \cup X^m$. We then retrain the unlearned model from scratch after receiving every $un$ unlearning requests of mitigation samples. Last, we evaluate the standard accuracy (ACC) and attack success rate (ASR) of initial and unlearned models based on the original testing dataset and the one with the trigger, respectively.

\begin{table}[t]\footnotesize
\centering
\caption{The effectiveness of Input-Targeted-based \methode.}
\label{tab:it_effective}
\begin{tabular}{|c||c|c||c|c||c|c||c|c|}
\hline
Dataset & \multicolumn{2}{c||}{MNIST} & \multicolumn{2}{c||}{FMNIST} & \multicolumn{2}{c||}{GTSRB} & \multicolumn{2}{c|}{CIFAR10} \\
\hline
Model & ACC & ASR & ACC & ASR & ACC & ASR & ACC & ASR \\
\hline
Initial & 98.77 & 0 & 90.57 & 0 & 96.61 & 0 & 87.08 & 0 \\
Unlearned & 98.83 & 100 & 90.13 & 100 & 96.47 & 100 & 87.23 & 100 \\
\hline
\end{tabular}
\end{table}

\noindent \textbf{Input-Targeted-based Attack.} For input-targeted attack, we first randomly select a seed image $(x, y)$ and a target label for poison data $y^p \neq y$. We apply uniform noise on the seed data to generate $|X^p|=5$ poison data and $|X^m|=15$ mitigation data (except for CIFAR10, which is set to be 10) independently. Since the quantity of mitigation data us small, we just forget then all at once to obtain the unlearned model. The attack success rate is evaluated on the original seed image and 50 generated backdoor samples that are orthogonal to the training set $X^p$ and $X^m$.

The experimental results are shown in Table~\ref{tab:it_effective}. Our observations reveal the following: 1) compared with the standard model, the accuracy loss of both the initial model and the unlearned model is negligible, i.e., with a maximum decrease of 0.34\%; 2) the attack success rate before and after unlearning all the mitigation samples is 0 and 100\% with only 15 (10 for CIFAR10) mitigation samples and 5 poison samples, respectively.

\begin{table*}[t]\small
\centering
\caption{The effectiveness of BadNets-based \methode.}
\label{tab:badnets_effective}
\begin{tabular}{|c||c|c||c|c||c|c||c|c|}
\hline
Dataset & \multicolumn{2}{c||}{MNIST} & \multicolumn{2}{c||}{FMNIST} & \multicolumn{2}{c||}{GTSRB} & \multicolumn{2}{c|}{CIFAR10} \\
\hline
$(\alpha, |X^p|, |X^m|)$ & \multicolumn{2}{c||}{(0.5, 50, 150)} & \multicolumn{2}{c||}{(0.5, 50, 100)} & \multicolumn{2}{c||}{(0.7, 300, 600)} & \multicolumn{2}{c|}{(0.7, 300, 100)} \\
\hline
RR & ACC & ASR & ACC & ASR & ACC & ASR & ACC & ASR \\
\hline
0 & 98.83 & 4.84 & 90.25 & 1.90 & 96.07 & 5.09 & 86.83 & 0.45 \\
0.1 & 98.76 & 5.51 & 90.38 & 2.10 & 96.36 & 9.99 & 86.95 & 0.76 \\
0.2 & 98.79 & 8.58 & 90.33 & 3.08 & 96.16 & 9.83 & 86.82 & 1.91 \\
0.3 & 98.70 & 12.73 & 90.39 & 2.60 & 96.19 & 13.90 & 87.02 & 1.74 \\
0.4 & 98.70 & 13.58 & 90.42 & 2.61 & 96.41 & 16.85 & 86.83 & 3.89 \\
0.5 & 98.84 & 16.83 & 90.37 & 6.56 & 96.23 & 22.07 & 86.95 & 5.24 \\
0.6 & 98.76 & 29.81 & 90.38 & 9.68 & 96.22 & 25.24 & 86.92 & 7.84 \\
0.7 & 98.83 & 37.51 & 90.14 & 14.94 & 96.36 & 29.76 & 86.89 & 20.27 \\
0.8 & 98.82 & 58.91 & 90.36 & 24.84 & 96.31 & 45.50 & 86.87 & 40.17 \\
0.9 & 98.73 & 81.25 & 90.20 & 44.47 & 96.31 & 58.87 & 86.93 & 57.87 \\
1.0 & 98.80 & 93.92 & 90.09 & 88.95 & 96.20 & 78.03 & 86.95 & 77.66 \\
\hline
\end{tabular}
\end{table*}

\noindent \textbf{BadNets-based Attack.} To evaluate the BadNets-based Attack, we randomly select $|X^p|$ benign images and embed a specific pattern, i.e., a $4 * 4$ (around 2\% of the entire image) white square,  with a hyper-parameter $\alpha$ at the bottom right of each of them to generate poison data. The target label is randomly chosen and assigned to each poison data. For mitigation samples, they are padded with the trigger in the same way, however labeled by the ground-truth classes. After the initial model is trained, we apply machine unlearning in a batch setting, i.e., we batch $10\% * |X^m|$ unlearning requests and retrain the model from scratch based on the left-over training set. Lastly, we generate a set of backdoor images by adding the pattern to each data in the clean testing dataset, whose label is not the target one, and collect the attack success rate on them.

The accuracy and attack success rate of initial model and unlearned models are represented in Table~\ref{tab:badnets_effective}. The first four columns show the experimental datasets and their corresponding settings, including $\alpha$, $|X^p|$ and $|X^m|$. The relationships among these three hyper-parameters are: 1) in order to achieve similar attack success rate of final unlearned model, the number of poison samples, $|X^p|$, should be increased when choosing a smaller $\alpha$; 2) when $|X^p|$ ($\alpha$) is fixed, the number of mitigation samples, $|X^m|$, is positively correlated with $\alpha$ ($|X^p|$) to ensure that the initial model is benign, i.e., with a small attack success rate. In the evaluation, we set different value of $\alpha$ for different scale datasets, e.g., 0.5 and 0.7 for grayscale and RGB images respectively, to balance the attack success rate and the number of poison and mitigation images (this setting is used in the following experiments). RR in the first column is the ratio of unlearning requests.

It can be observed that for RGB (grayscale) datasets, when $\alpha$ and the number of poison samples are fixed, the more complex the dataset is, the larger the number of mitigation images needed. For instance, for GTSRB, 600 mitigation samples are needed, whereas for CIFAR10, 100 images suffice. Similarly, for MNIST and FMNIST, the numbers are 150 and 100 respectively. And for all the datasets, the number of poison and mitigation samples are less than 1\% and 2\% of the size of the clean training dataset, respectively. Further, as the number of unlearned samples increases, the attack success rate initially increases at a slow rate, i.e., from the initial model to unlearning half of the mitigation data, the ASR of the model increased by only 9.61\% on average, then rapidly increases (refer to the blue solid line in Figure~\ref{fig:badnets_sisa}), i.e., after forgetting the remaining 1/2 mitigation data, the model's ASR increased by 71.97\% on average. In addition, the accuracy is comparable for the model trained with poison and mitigation data and the one trained with only clean training data, on average, the loss of model accuracy is only 0.15\% for all the experimental datasets.

\begin{framed}
\noindent Answer to RQ1: \method can successfully embed a backdoor into the model through machine unlearning, i.e., with a low ASR of the initial model and a moderately high ASR of the unlearned model. For Input-Targeted-based Attack, it can be achieved with 5 poison samples and no more than 15 mitigation samples; for BadNets-based Attack, the number of poison and mitigation samples are less than 1\% and 2\% of the size of the clean training dataset, respectively.
\end{framed}

\subsection{RQ2: Against SISA}
\label{subsec:rq2}

\begin{table*}[t]\small
\centering
\caption{Experiment setting of \method w.r.t different number of shards ($R$ = 1).}
\label{tab:sisa_setting}
\begin{tabular}{|c|c||c|c|c||c|c|c||c|c|c||c|c|c|}
\hline
Dataset & Str. & \multicolumn{3}{c||}{S = 3} & \multicolumn{3}{c||}{S = 5} & \multicolumn{3}{c||}{S = 7} & \multicolumn{3}{c|}{S = 9} \\
\cline{3-14}
& & ACC & $|X^p|$ & $|X^m|$ & ACC & $|X^p|$ & $|X^m|$ & ACC & $|X^p|$ & $|X^m|$ & ACC & $|X^p|$ & $|X^m|$ \\ 
\hline
\multirow{2}{*}{MNIST} & IT & \multirow{2}{*}{98.62} & 20 & 50 & \multirow{2}{*}{98.32} & 30 & 60 & \multirow{2}{*}{98.03} & 40 & 60 & \multirow{2}{*}{97.76} & 60 & 70 \\
& BN & & 150 & 450 & & 200 & 600 & & 300 & 800 & & 400 & 1100 \\
\hline
\multirow{2}{*}{FMNIST} & IT & \multirow{2}{*}{90.46} & 10 & 30 & \multirow{2}{*}{89.77} & 10 & 30 & \multirow{2}{*}{89.34} & 15 & 40 & \multirow{2}{*}{89.18} & 20 & 40 \\
& BN & & 150 & 200 & & 200 & 300 & & 250 & 300 & & 300 & 400 \\
\hline
\multirow{2}{*}{GTSRB} & IT & \multirow{2}{*}{96.32} & 5 & 15 & \multirow{2}{*}{95.76} & 10 & 30 & \multirow{2}{*}{95.60} & 10 & 30 & \multirow{2}{*}{94.96} & 15 & 30 \\
& BN & & 600 & 1100 & & 1000 & 2000 & & 1300 & 2500 & & 1500 & 2800 \\
\hline
\multirow{2}{*}{CIFAR10} & IT & \multirow{2}{*}{83.20} & 5 & 10 & \multirow{2}{*}{81.29} & 5 & 15 & \multirow{2}{*}{79.83} & 10 & 15 & \multirow{2}{*}{78.36} & 10 & 20 \\
& BN & & 700 & 300 & & 1000 & 600 & & 1200 & 700 & & 1500 & 1000 \\
\hline
\end{tabular}
\end{table*}

\begin{figure}[t]
\centering
\subfigure[MNIST]{
\includegraphics[width=0.22\textwidth]{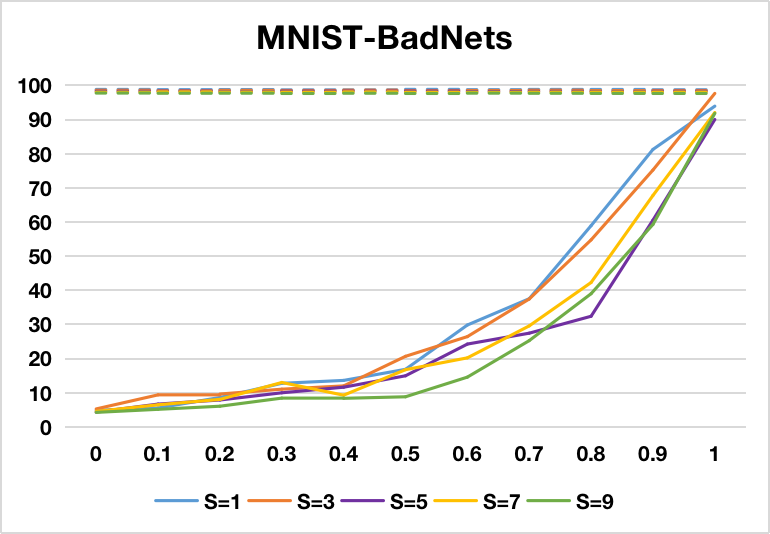}
\label{fig:mnist-badnets}
}
\subfigure[FMNIST]{
\includegraphics[width=0.22\textwidth]{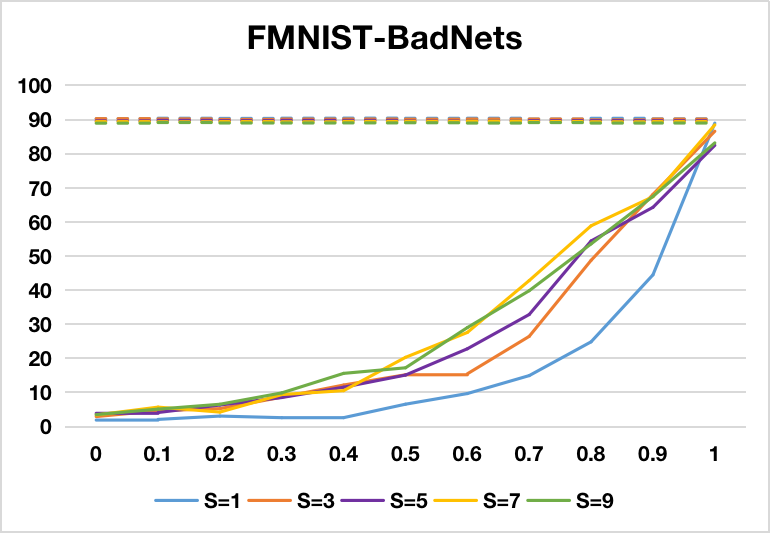}
\label{fig:fmnist-badnets}
}
\subfigure[GTSRB]{
\includegraphics[width=0.22\textwidth]{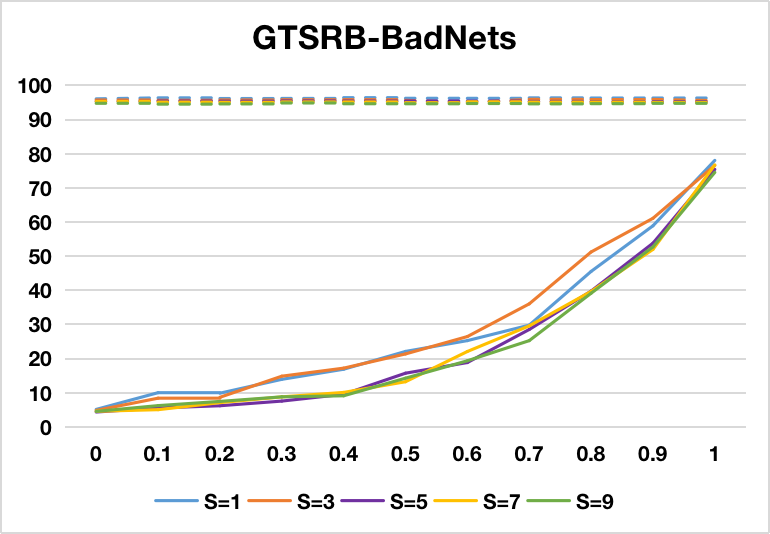}
\label{fig:gtsrb-badnets}
}
\subfigure[CIFAR10]{
\includegraphics[width=0.22\textwidth]{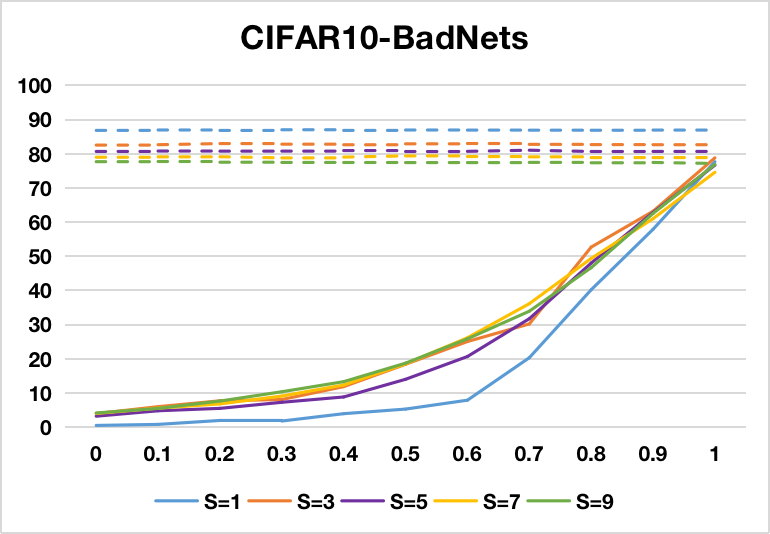}
\label{fig:cifar10-badnets}
}
\caption{The effectiveness of BadNets-based \method w.r.t different number of shards ($R$=1).}
\label{fig:badnets_sisa}
\end{figure}

Next, we evaluate the effectiveness of \method under different settings of SISA~\cite{sisa}, a state-of-the-art machine unlearning framework. Note that the unlearning strategy in Section~\ref{subsec:rq1} can be regarded as SISA training with $(S, R) = (1, 1)$.

For evaluation, we utilize the same method as in Section~\ref{subsec:rq1} to generate the poison and mitigation samples and randomly select the target label. Then the whole training dataset $X = X^c \cup X^p \cup X^m$ is randomly divided into $S$ shards and $R$ slices. We conduct a comprehensive experiment to evaluate the impact of sharding and slicing on the effectiveness of \methode.

\noindent \textbf{Impact of Sharding.} To evaluate the effect of sharding, we fix the number of slices as 1, and choose the number of shards from $[1,3,5,7,9]$. We show the settings of our attack algorithm in Table~\ref{tab:sisa_setting}, including the standard testing accuracy and the number of poison and mitigation samples. For Input-Targeted-based \methode, the attack success rate of the initial model and the final unlearned model are 0 and 100\%, respectively. The results of BadNets-based \method is represented in Figure~\ref{fig:badnets_sisa}, where the x-axis and y-axis are the ratio of unlearning requests and the accuracy; the solid and dotted line represents the attack success rate and accuracy, respectively. 

After analyzing the data presented in Table~\ref{tab:sisa_setting} and Figure~\ref{fig:badnets_sisa}, several observations can be made. First, increasing the number of sub-datasets used for SISA training will reduce the accuracy of the model, with the extent of the drop depending on the complexity of the dataset. That is, the more complex the dataset is, the more accuracy on clean samples drops. For instance, when the number of sub-datasets was increased from 1 to 9, the standard accuracy of MNIST only dropped by 0.92\%, while the standard accuracy of CIFAR10 decreased by 8.85\%. Second, for both Input-Targeted-based and BadNets-based attacks, to achieve a similar level of attack effectiveness, the number of poison and mitigation samples required increases almost linearly with the number of shards used. This is probably because sub-models trained on partial datasets have similar accuracy to the model trained on the whole dataset, i.e., they have the same generalization ability, and therefore require an equivalent amount of poison and mitigation samples to achieve the same attack target. Based on these observations, dividing the dataset into multiple subsets for training sub-models and then ensembling the predictions can be considered a potential backdoor attack defense strategy. However, the number of shards used needs to be carefully designed to balance the prediction accuracy and the resources of the whole model, such as the storage space required, which is proportional to the number of shards used, i.e., $\Theta(S)$.

\begin{figure*}[t]
\centering
\subfigure[MNIST]{
\includegraphics[width=0.23\textwidth]{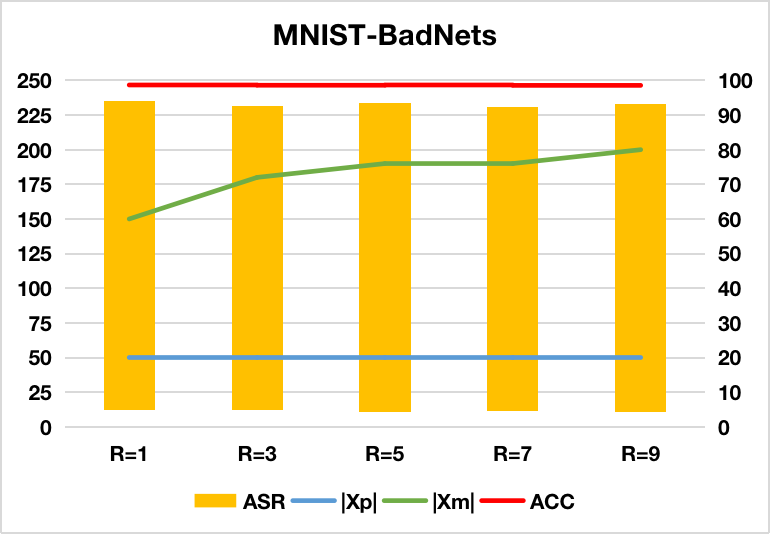}
\label{fig:mnist-slice}
}
\subfigure[FMNIST]{
\includegraphics[width=0.23\textwidth]{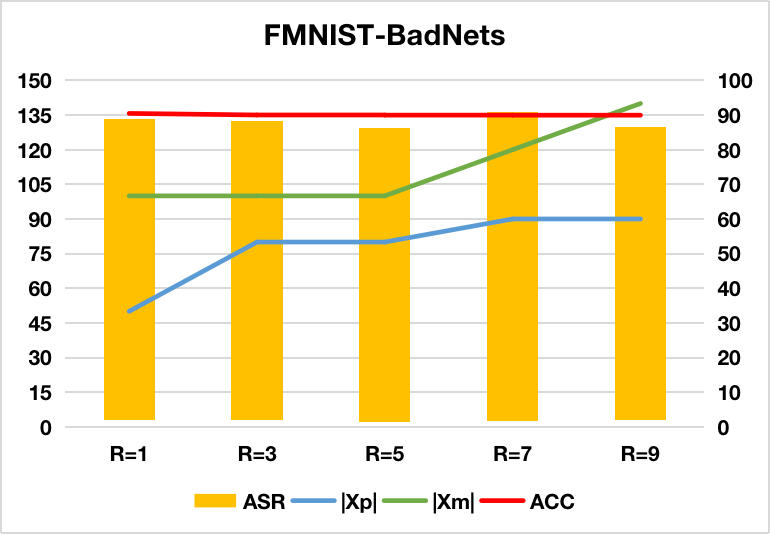}
\label{fig:fmnist-slice}
}
\subfigure[GTSRB]{
\includegraphics[width=0.23\textwidth]{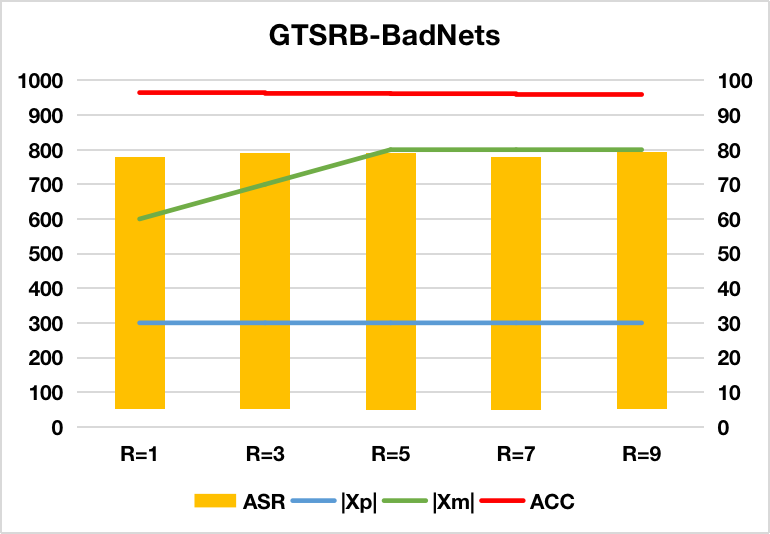}
\label{fig:gtsrb-slice}
}
\subfigure[CIFAR10]{
\includegraphics[width=0.23\textwidth]{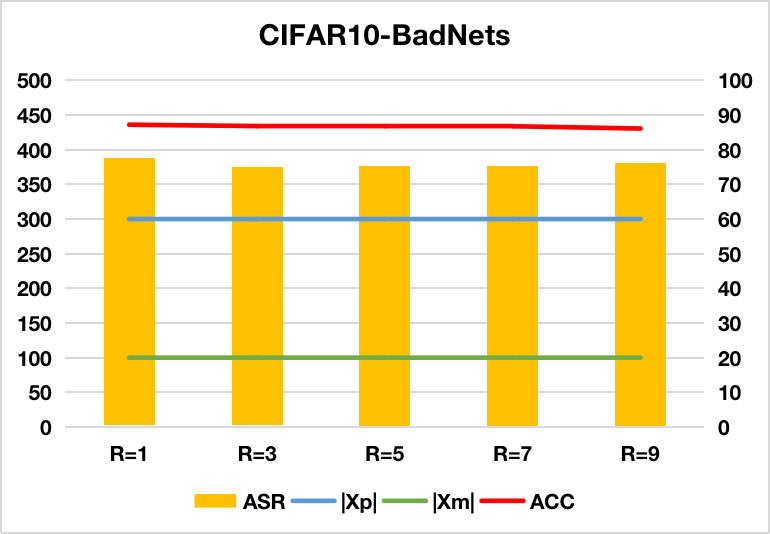}
\label{fig:cifar10-slice}
}
\caption{The effectiveness of BadNets-based \method w.r.t different number of slices ($S$=1).}
\label{fig:badnets_slice}
\end{figure*}

\begin{figure*}[t]
\centering
\subfigure[MNIST]{
\includegraphics[width=0.23\textwidth]{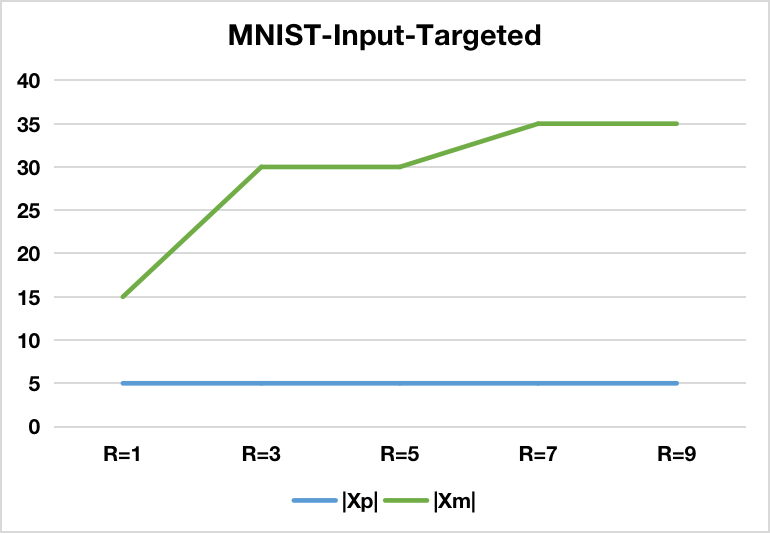}
\label{fig:mnist-it-slice}
}
\subfigure[FMNIST]{
\includegraphics[width=0.23\textwidth]{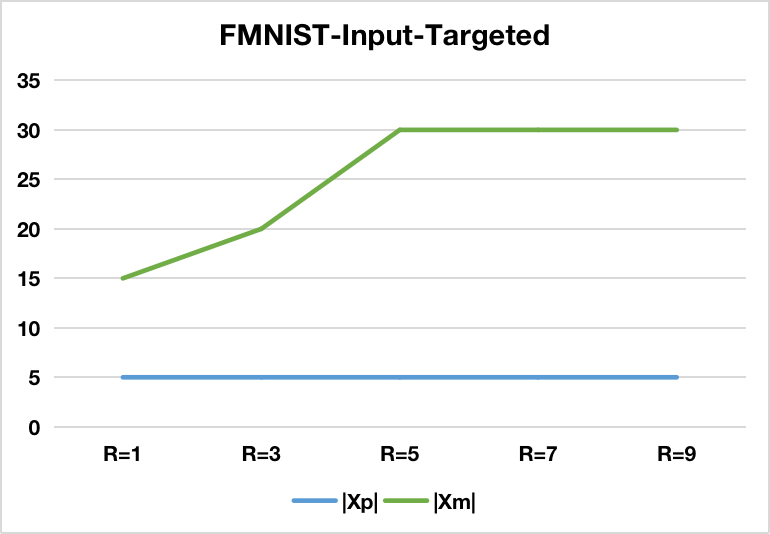}
\label{fig:fmnist-it-slice}
}
\subfigure[GTSRB]{
\includegraphics[width=0.23\textwidth]{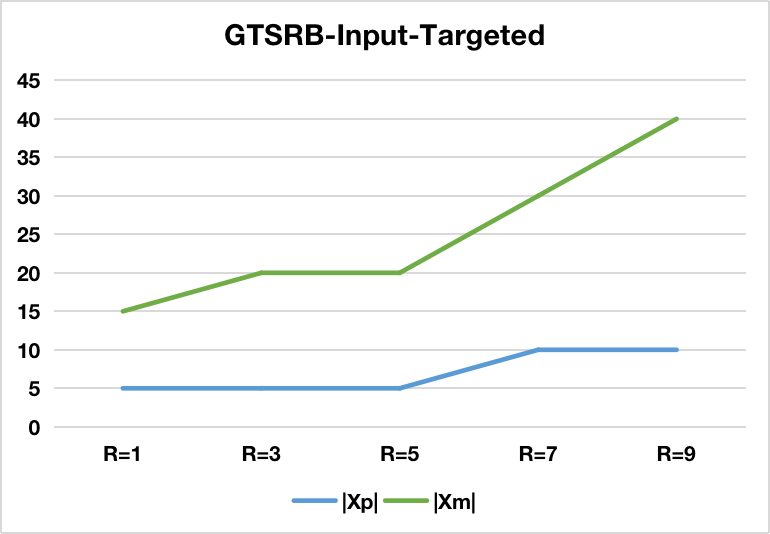}
\label{fig:gtsrb-it-slice}
}
\subfigure[CIFAR10]{
\includegraphics[width=0.23\textwidth]{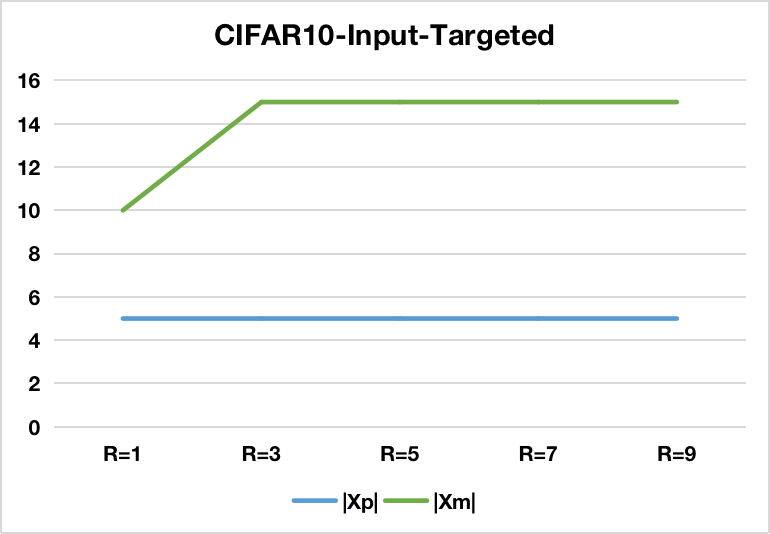}
\label{fig:cifar10-it-slice}
}
\caption{The effectiveness of Input-Targeted-based \method w.r.t different number of slices ($S$=1).}
\label{fig:it_slice}
\end{figure*}

\noindent \textbf{Impact of Slicing.} To evaluate the effect of slicing, the number of shards is fixed as 1, and the number of slices $R$ is selected from $[1,3,5,7,9]$. For both BadNets-based and Input-Targeted-based \methode, we adjust the number of poison and mitigation samples to make the attack success rate of the initial models and final unlearned models reach a similar level, respectively. The settings and the results of BadNets-based and Input-Targeted-based \method are shown in Figure~\ref{fig:badnets_slice} and ~\ref{fig:it_slice}, respectively. The x-axis is the number of slices, and the left and right y-axis are the amount of samples and the accuracy (or the attack success rate), respectively. The red line indicates the standard accuracy, the blue line represents the quantity of poison data, and the green line shows the amount of mitigation samples. In addition, for BadNets-based \methode, the lower and upper edges of the yellow histogram respectively represent the attack success rate of the initial model and the final unlearned model (after unlearning all the mitigation samples). For Input-Targeted-based \methode, the attack success rate is 0 and 100\% for models before and after unlearning all the mitigation data, respectively.

From Figure~\ref{fig:badnets_slice} and ~\ref{fig:it_slice}, it can be observed that: 1) compare to the approach without slicing, slicing has almost no impact on the standard accuracy of the model. Among all slicing settings, their standard accuracy decreased by only 0.37\% on average. 2) As the number of slices increases, the threshold of poison samples necessary to launch a successful attack on the majority of datasets remains stable, with only a slight uptick for some cases. Specifically, for both BadNets-based and Input-Targeted-based attacks, 3/4 of datasets require the same amount of poison samples, while only one dataset experiences a mild increase. For instance, on FMNIST, the threshold for BadNets-based attack has risen by 10, while for Input-Targeted-based attack, the increase is 5 on GTSRB. 3) The number of mitigation samples slowly increases as the number of slices increases. When the number of slices increases by 8 times (from R=1 to R=9), the required number of mitigation samples for BadNets-based and Input-Targeted-based attacks increases by an average of 26.7\% and 112.5\%, respectively. Overall, increasing the number of slices does not significantly enhance the difficulty of \methode, however, it also raise the additional storage overhead of the models which is $\Theta(R)$.

\begin{framed}
\noindent Answer to RQ2: SISA will increase the difficulty of infecting the model with a backdoor by \methode. As the number of shards increases, the amount of poison and mitigation samples will also increase, however, it will also lead to the reduction of model accuracy and the raise of storage resources. In addition, slicing has a minor impact on both the standard accuracy and the difficulty of unlearning-based backdoor attacks.
\end{framed}

\subsection{RQ3: Evasion}
\label{subsec:rq3}
From Table~\ref{tab:badnets_effective} and Figure~\ref{fig:badnets_sisa}, it can be observed that for BadNets-based attack, the attack success rate of the initial model on MNIST and GTSRB is about 5\%. Due to the sample construction method we employed, we are limited in our ability to further reduce the attack success rate with mitigation samples. As a result, the number of mitigation samples required to achieve additional reductions will increase significantly, making subsequent steps in machine unlearning more difficult. In the following, we adopt to state-of-the-art backdoor detection algorithms to evaluate whether the backdoor in the initial model can be detected.
\begin{itemize}
\item Neural Cleanse~\cite{neural_cleanse}. Based on the hypothesis that the backdoor model establishes a "shortcut" (trigger) from the area in the space assigned to the label to the area assigned to the target label, it first reverses the trigger by solving a joint optimization objective of maximizing the misclassification probability on the samples with the trigger and minimizing the size of the trigger, and then calculate the anomaly index of the $L_1$ norm of the trigger based on the Median Absolute Deviation (MAD). The anomaly index exceeding 2 suggests that the corresponding trigger has more than 95\% likelihood of being an outlier, implying the presence of a backdoor in the model.
\item Randomized Channel Shuffling~\cite{rcs}. It hypothesizes that the trigger features are represented sparsely in only a limited number of filters, whereas the features of clean data are encoded across a larger amount of filters. Therefore, it randomly shuffles the order of filters and calculate the output shift of the last convolutional layer based on the original model. It also uses Median Absolute Deviation to detect the outlier of the standard deviation of representation shift, i.e., the backdoor model, by properly setting the threshold.
\end{itemize}

\begin{table*}[t]\small
\centering
\caption{The Evasion Performance of Initial Model.}
\label{tab:evasion}
\begin{tabular}{|c|c|c|c|c|c|c|c|c|c|}
\hline
Dataset & \multicolumn{2}{c|}{MNIST} & \multicolumn{2}{c|}{FMNIST} & \multicolumn{2}{c|}{GTSRB} & \multicolumn{2}{c|}{CIFAR10}\\
\cline{2-9}
(S,R) & AI & AUROC & AI & AUROC & AI & AUROC & AI & AUROC \\
\hline
(1,1) & 1.080 & 0.373 & 0.894 & 0.552 & 1.150 & 0.613 & 1.591 & 0.582 \\
\hline
(3,1) & 1.172 & 0.444 & 0.977 & 0.413 & 0.948 & 0.616 & 1.117 & 0.531 \\
\hline
(5,1) & 1.144 & 0.466 & 0.783 & 0.517 & 0.667 & 0.607 & 1.057 & 0.453 \\
\hline
(7,1) & 0.841 & 0.466 & 0.790 & 0.533 & 0.703 & 0.582 & 1.078 & 0.392 \\
\hline
(9,1) & 0.786 & 0.398 & 1.012 & 0.413 & 0.982 & 0.511 & 0.897 & 0.496 \\
\hline
(1,3) & 1.031 & 0.392 & 1.162 & 0.544 & 0.858 & 0.608 & 1.866 & 0.256 \\
\hline
(1,5) & 0.930 & 0.387 & 0.939 & 0.416 & 1.003 & 0.464 & 1.691 & 0.373 \\
\hline
(1,7) & 0.962 & 0.456 & 1.487 & 0.584 & 1.240 & 0.384 & 1.605 & 0.360 \\
\hline
(1,9) & 0.959 & 0.560 & 0.837 & 0.520 & 0.641 & 0.392 & 1.350 & 0.227 \\
\hline
\end{tabular}
\end{table*}

We obtained the implementation of Neural Cleanse\footnote{https://github.com/bolunwang/backdoor} and Randomized Channel Shuffling\footnote{https://github.com/VITA-Group/Random-Shuffling-BackdoorDetect} from GitHub. To evaluate the evasion performance of Neural Cleanse, for each sub-model, we independently reverse the potential trigger for each label and then calculate the anomaly index of the target label. Since the reverse engineering starts from a random trigger pattern and mask, we ran Neural Cleanse 5 times on each model and reported the average results. In addition, for Randomized Channel Shuffling, we train 25 initial (backdoor) sub-models for each combination of different numbers of shards and slices, along with 25 benign models~\cite{rcs}. To improve diversity, each benign model is trained on 90\% randomly chosen clean training samples with random initial weights. We shuffle the channel order of only the last four convolutional layers (unless the model has fewer than four convolutional layers, in which case all convolutional layers are considered) to obtain the feature sensitivity curve. We measure the effectiveness of Randomized Channel Shuffling in detecting initial models utilizing the Area under Receiver Operating Characteristic Curve (AUROC)~\cite{mutation,rcs}. The closer the AUROC is to 1, the better the detection strategy performs.

The results are shown in Table~\ref{tab:evasion}. Column AI presents the average anomaly index with repect to the target label among all the sub-models for Neural Cleanse, and Column AUROC shows the detection performance for Randomized Channel Shuffling. It can be observed that on average: 1) the anomaly index for Neural Cleanse is only 1.062, and 2) the AUROC value for Randomized Channel Shuffling can only reach 0.469, which indicates that utilizing Randomized Channel Shuffling to distinguish benign models and initial models is worse than random guessing. One possible explanation is that the vast majority of existing detection algorithms are generally less sensitive to attacks with low success rates, as they require a relatively high attack success rate for accurate detection of backdoors. That is, if the backdoor attack is designed by the adversary in such a way that it does not significantly affect the model's performance, e.g., initial model in \methode, then it becomes difficult for the detection algorithm to identify the presence of a backdoor.

\begin{framed}
\noindent Answer to RQ3: Existing detection methods are inadequate for detecting backdoors in the model. Neural Cleanse has an average anomaly index of only about half the specified threshold (1.062/2), and the performance of Randomized Channel Shuffling is worse than random guessing.
\end{framed}

\section{Sample-based Defense}
\label{sec:defense}

In the previous sections, we have shown that the proposed unlearning-based backdoor attack, \methode, is effective. However, if we conduct a deep analysis of the backdoor implantation process, we can find that the backdoor attack strategies based on machine unlearning will gradually degenerate into the corresponding classical methods. Therefore, most of the existing defense methods, e.g., Neural Cleanse~\cite{neural_cleanse} and Randomized Channel Shuffling~\cite{rcs}, still can be used to defend against \methode. The practical difficulty is that in order to detect \methode, existing detection methods such as Neural Cleanse must be applied after every unlearning request, which would incur a lot of computational overhead. In addition, there are also some limitations of existing backdoor detection algorithms. First, they require prior knowledge of the trigger of the backdoor to detect it. If the adversary uses new or unknown triggers to implement the backdoor attack, they will not be able to detect it. Second, the backdoor detection algorithm based on reverse engineering relies on processing and analyzing a large amount of data, thus requiring high computational resources to support it, especially when dealing with large-scale datasets with numerous classes.

In the following, we introduce two dedicated methods for defensing \methode, which are only based on the model's output of the unlearning request. The complete defense pipeline involves first using the proposed methods to determine whether the unlearning request is malicious. If it is, the victim can either directly reject the request to forget the samples or apply the aforementioned defense algorithms to detect the backdoor after unlearning. It is worth noting that our detection algorithms occurs before model retraining, thus it only consumes minimal computational resources. Additionally, this approach does not require any prior knowledge about backdoor attacks, including triggers and target classes.

\subsection{Model Uncertainty}

We first try to determine whether the unlearning request from users is malicious based on the intuition that: 1) a well-trained model can perfectly fit its clean training data, meaning that it can correctly classify these samples with high confidence; 2) the mitigation samples are designed to minimize the impact of poison data on model predictions during training, and vice versa. Therefore, the two sets of samples constrain each other, making it challenging for the model to learn the mitigation samples effectively, resulting in a large loss value on the mitigation data.

We use the Gini impurity of the model output probability to measure the likelihood of the input being misclassified~\cite{deepgini,gini}, which is formally defined as follows.
\begin{equation}
Impurity = 1-\sum_{c=1}^C P(x)_c^2
\end{equation}
where $C$ is the total number of classes and $P$ denotes the output probability vector of initial model $D$. The more higher the impurity value, the more likely the input is to be an incorrectly classified (malicious) sample.

\subsection{Sub-Model Similarity}

The detection algorithm based on Gini impurity performs well when the output probability of the model on the mitigation sample is small. If the mitigation samples are carefully designed and have a significant impact on the prediction of the model trained with them, it becomes difficult for the algorithm to differentiate between the mitigation samples and the clean samples. To address this challenge, we propose an alternative detection method under the SISA setting that utilizes the standard deviation among multiple predictions of sub-models to identify malicious unlearning requests. The proposed detection method is based on the intuition that the presence of poison samples can lead to poor generalization ability of mitigation data. This means that mitigation samples are more likely to obtain high prediction confidence only on the corresponding sub-model trained using them, resulting in a larger prediction variation among different sub-models. On the other hand, clean samples are expected to have more consistent predictions across well-trained sub-models due to their generalization ability, resulting in a smaller standard deviation. The similarity of sub-models is formally defined as follows.

\begin{equation}
Similarity = Std(\bigcup_{s=1}^{S}P_s(x)_{c})
\end{equation}
where $P_s(\cdot)$ denotes the output probability vector of s-th sub-model and c is the ground-truth label. A smaller similarity indicates that the sample is more likely to be clean.

When an unlearning request is made, we first calculate either the Gini impurity of the current model or the output standard deviation among all the sub-models. If the resulting value is below a certain threshold, we accept the request and remove the sample from the model. However, if the result is above the threshold, we identify the sample as a mitigation sample and reject the unlearning request.
\section{Evaluation of Defense}
\label{sec:experiment_defense}

In this section, we design experiments that focus on answering the following research question:
\begin{itemize}
\item RQ4: Can our proposed detection strategies effectively identify malicious unlearning requests? 
\end{itemize}


To assess the effectiveness of our proposed detection algorithms, i.e., Model Uncertainty and Sub-Model Similarity, we randomly select an equal number of clean and mitigation samples as potential unlearning requests submitted by the attacker, and then calculate the corresponding AUROC score. A higher AUROC indicates a stronger ability to distinguish between the clean and mitigation sample.

\begin{table*}[t]\small
\centering
\caption{AUROC results for Model Uncertainty against Input-Targeted-based \methode.}
\label{tab:mu-it}
\begin{tabular}{|c||c|c|c||c|c|c||c|c|c||c|c|c|}
\hline
RR & \multicolumn{3}{c||}{MNIST} & \multicolumn{3}{c||}{FMNIST} & \multicolumn{3}{c||}{GTSRB} & \multicolumn{3}{c|}{CIFAR10} \\
\cline{2-13}
& (1,1) & (5, 1) & (1, 5) & (1,1) & (5, 1) & (1, 5) & (1,1) & (5, 1) & (1, 5) & (1,1) & (5, 1) & (1, 5) \\
\hline
0 & 0.947 & 0.964 & 0.958 & 0.909 & 0.877 & 0.879 & 1.000 & 0.977 & 1.000 & 1.000 & 0.840 & 0.998 \\
0.2 & 0.983 & 0.969 & 0.986 & 0.905 & 0.859 & 0.891 & 0.991 & 0.988 & 1.000 & 1.000 & 0.855 & 0.998 \\
0.4 & 1.000 & 0.974 & 0.989 & 0.865 & 0.857 & 0.892 & 1.000 & 0.983 & 0.992 & 1.000 & 0.872 & 0.998 \\
0.6 & 1.000 & 0.973 & 1.000 & 0.950 & 0.829 & 0.902 & 1.000 & 0.990 & 1.000 & 1.000 & 0.901 & 1.000 \\
0.8 & 1.000 & 0.978 & 0.967 & 0.910 & 0.907 & 0.944 & 1.000 & 0.995 & 1.000 & 1.000 & 0.843 & 1.000 \\
\hline
\end{tabular}
\end{table*}

\begin{table*}[t]\small
\centering
\caption{AUROC results for Model Uncertainty against BadNets-based \methode.}
\label{tab:mu-bn}
\begin{tabular}{|c||c|c|c||c|c|c||c|c|c||c|c|c|}
\hline
RR & \multicolumn{3}{c||}{MNIST} & \multicolumn{3}{c||}{FMNIST} & \multicolumn{3}{c||}{GTSRB} & \multicolumn{3}{c|}{CIFAR10} \\
\cline{2-13}
& (1,1) & (5, 1) & (1, 5) & (1,1) & (5, 1) & (1, 5) & (1,1) & (5, 1) & (1, 5) & (1,1) & (5, 1) & (1, 5) \\
\hline
0 & 0.836 & 0.712 & 0.831 & 0.767 & 0.652 & 0.733 & 0.836 & 0.645 & 0.865 & 0.919 & 0.789 & 0.900 \\
0.2 & 0.872 & 0.757 & 0.821 & 0.767 & 0.655 & 0.731 & 0.861 & 0.672 & 0.869 & 0.920 & 0.794 & 0.963 \\
0.4 & 0.886 & 0.798 & 0.826 & 0.771 & 0.662 & 0.791 & 0.869 & 0.689 & 0.890 & 0.949 & 0.804 & 0.953 \\
0.6 & 0.912 & 0.832 & 0.903 & 0.807 & 0.685 & 0.826 & 0.880 & 0.718 & 0.890 & 0.958 & 0.824 & 0.970 \\
0.8 & 0.962 & 0.843 & 0.937 & 0.851 & 0.744 & 0.844 & 0.875 & 0.732 & 0.885 & 0.806 & 0.738 & 0.929 \\
\hline
\end{tabular}
\end{table*}

\noindent \textbf{Model Uncertainty.} For the detection method based on model uncertainty, we conduct the experiment under three combinations of shards and slices, i.e., $(S,R)=(1,1),(5,1),(1,5)$. It is worth noting that when the number of shards is greater than 1, i.e., there are multiple sub-models, we sample the clean data from each sub-training dataset independently and compute the Gini impurity based on the respective sub-model. 

\begin{figure}[t]
\centering
\subfigure[FMNIST (Mitigation Sample)]{
\includegraphics[width=0.225\textwidth]{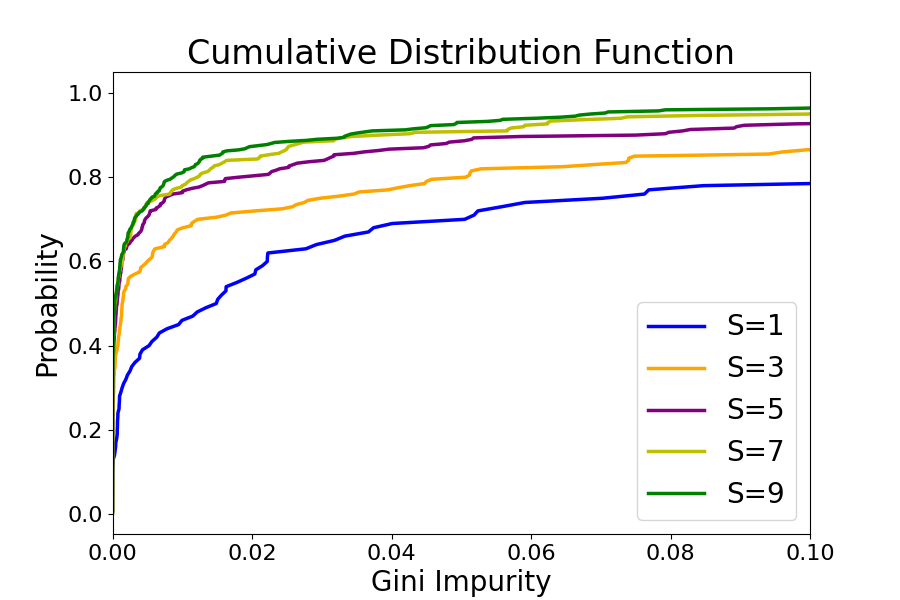}
\label{fig:fmnist-mu}
}
\subfigure[CIFAR10 (Clean Sample)]{
\includegraphics[width=0.225\textwidth]{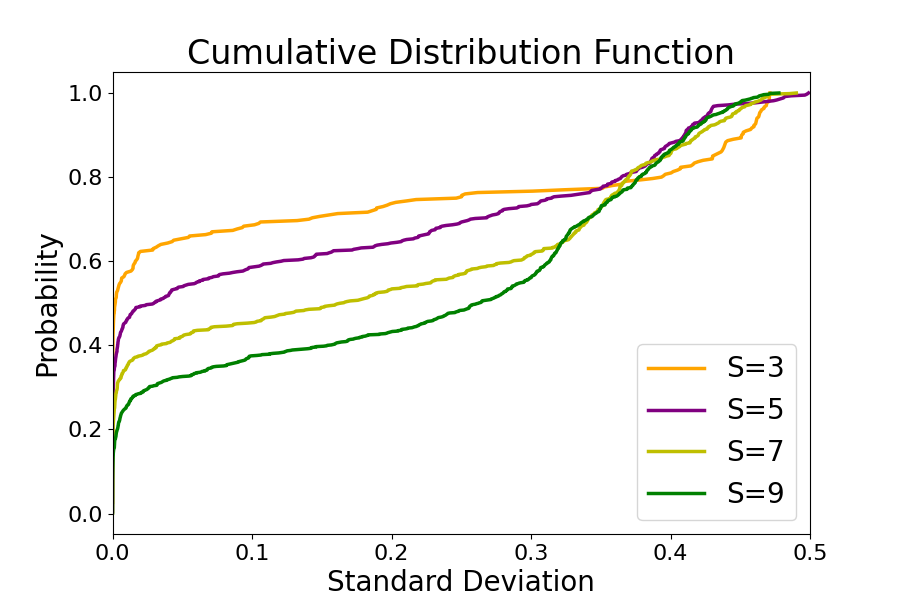}
\label{fig:cifar10-sms}
}
\caption{The Cumulative Distribution Function (CDF) w.r.t Gini impurity and standard deviation for BadNets-based \methode.}
\label{fig:cdf_sisa}
\end{figure}

Table~\ref{tab:mu-it} and ~\ref{tab:mu-bn} present the AUROC results on models unlearning each $20\% * |X^m|$ requests against Input-Targeted-based and BadNets-based \methode. We can observe that when $(S,R)=(1,1)$, whether against Input-Targeted-based or BadNets-based \methode, Gini impurity can effectively identify the mitigation and clean samples, especially for the Input-Targeted-based attack, the AUROC results reaches 1 (i.e., a perfect classifier) in over half the cases (12/20). We can also observe that increasing the number of shards will lead to a significant decrease in the AUROC value, especially for BadNets-based attack, the AUROC value of the initial model decreases by 0.145 on average when the number of shards is increased from 1 to 5. As shown in Figure~\ref{fig:fmnist-mu}, one possible explanation for this observation is that, under the same training configuration, the sub-model can better fit mitigation samples as the number of shards increases. This is due to the fact that an increase in the number of shards reduces the number of samples in the sub-model's training set, but the number of mitigation data remains almost unchanged (as demonstrated in Section~\ref{subsec:rq2}). Furthermore, the performance of Gini impurity in detecting mitigation samples is not adversely affected by slicing, when compared to the approach without slicing, on average, the AUROC values only decrease by 0.003 for both Input-Targeted-based and BadNets-based attack, respectively.

\begin{table*}[t]
\centering
\caption{AUROC results for Sub-Model Similarity against BadNets-based \methode.}
\label{tab:sms-bn}
\begin{tabular}{|c||c|c|c|c||c|c|c|c||c|c|c|c||c|c|c|c|}
\hline
RR & \multicolumn{4}{c||}{MNIST} & \multicolumn{4}{c||}{FMNIST} & \multicolumn{4}{c||}{GTSRB} & \multicolumn{4}{c|}{CIFAR10} \\
\cline{2-17}
& (3,1) & (5, 1) & (7, 1) & (9,1) & (3,1) & (5, 1) & (7, 1) & (9,1) & (3,1) & (5, 1) & (7, 1) & (9,1) & (3,1) & (5, 1) & (7, 1) & (9,1) \\
\hline
0 & 0.883 & 0.853 & 0.858 & 0.829 & 0.858 & 0.837 & 0.829 & 0.813 & 0.809 & 0.763 & 0.732 & 0.698 & 0.856 & 0.820 & 0.793 & 0.755 \\
0.2 & 0.918 & 0.878 & 0.860 & 0.863 & 0.870 & 0.850 & 0.852 & 0.843 & 0.815 & 0.774 & 0.757 & 0.735 & 0.864 & 0.833 & 0.795 & 0.769 \\
0.4 & 0.933 & 0.877 & 0.905 & 0.879 & 0.872 & 0.871 & 0.875 & 0.885 & 0.833 & 0.781 & 0.792 & 0.753 & 0.865 & 0.841 & 0.803 & 0.769 \\
0.6 & 0.942 & 0.932 & 0.938 & 0.894 & 0.882 & 0.909 & 0.897 & 0.891 & 0.836 & 0.801 & 0.844 & 0.805 & 0.876 & 0.845 & 0.805 & 0.757 \\
0.8 & 0.986 & 0.968 & 0.965 & 0.961 & 0.914 & 0.935 & 0.922 & 0.919 & 0.870 & 0.840 & 0.871 & 0.854 & 0.801 & 0.789 & 0.775 & 0.727 \\
\hline
\end{tabular}
\end{table*}

\noindent \textbf{Sub-Model Similarity.} Since the uncertainty-based method has been shown to be effective in defending against Input-Targeted-based attack, which indicates low output probability for mitigation sample on the corresponding sub-model, it suggests that the defense method based on sub-model similarity will have poor performance in this case. As a result, we only evaluated the detection effectiveness of similarity-based algorithm against BadNets-based attack. The experiment is conducted with varying numbers of shards, while keeping the number of slices equal to 1.

The result is shown in Table~\ref{tab:sms-bn}. It can be observed that our similarity-based detection algorithm is effectively in identifying unlearning mitigation samples against BadNets-based \methode, the AUROC value for initial models is 0.812 on average across all experiment settings. Combined with the results shown in Table~\ref{tab:mu-bn}, the results demonstrate that the detection method based on the standard deviation of outputs from multiple sub-models outperforms the Gini impurity-based method that relies on a single model's output when SISA divides the dataset into multiple shards. For example, the experimental results showed an average improvement of 0.113 when $S=5$. Additionally, increasing the number of shards can also lead to a decrease in the detection performance of Sub-Model Similarity. This can be attributed to the fact that as the number of training samples per sub-model decreases, the generalization ability of the sub-model weakens, resulting in increasingly unstable outputs for clean samples across different sub-models, as shown in Figure~\ref{fig:cifar10-sms}.

\begin{framed}
\noindent Answer to RQ4: Our proposed sample-based detection algorithms have shown to be effective in identifying malicious unlearning requests of mitigation samples. However, their performances are significantly reduced when sharding is applied.
\end{framed}

\section{Related Work}
\label{sec:relatedwork}

\noindent \textbf{Poisoning-based Backdoor Attack.} Backdoor attacks on machine learning models have been an active area of research in recent years. 
In this work, we focus on the problem of data-poisoning attacks involving modifying the training data to add a backdoor to the model. Gu et al.~\cite{badnets} were the first to introduce the concept of backdoor attacks in deep learning models and outline the fundamental steps of a backdoor attack. It involves adding trigger (e.g., a pattern of bright pixels or an image of bomb) as poisoned data to normal data, labeling the poisoned data with the target label specified by the attacker, and training the poisoned and normal data together. Liu et al.~\cite{trojaning} introduced the concept of using reverse engineering to apply backdoor attack, where triggers are obtained through optimization based on the activation value of neuron and poison samples are constructed when real dataset cannot be accessed. Chen et al.~\cite{targeted_backdoor} were pioneers in exploring invisible backdoor attacks, which overlayed the trigger to some extent onto the original image pixels, instead of replaced directly. They employed two backdoor strategies: the Input-instance-key mode utilized random noise as the trigger and the Pattern-key mode mixed the original image with a trigger based on a specified pattern to a certain extent. Later, more works dedicated to this research have been proposed~\cite{invisible1,invisible2,invisible3,invisible4}, e.g., Li et al.~\cite{invisible1} embedded the trigger into the bit space by image steganography techniques and Doan et al.~\cite{invisible3} suggested regularizing the $L_{p}$ norm of the perturbation during optimization of the backdoor trigger. Instead of modifing the image digitally with non-semantic, Bagdasaryan et al. ~\cite{semantic} first proposed semantic backdoor attacks~\cite{semantic1,semantic2}, which takes a semantic aspect of data as the trigger, e.g., cars with racing stripes or plaid shirt. Therefore, the infected model will automatically misclassify images that contain the pre-defined semantic information without any digital alteration. The above methods achieve the backdoor attack by directly training on the clean and poison samples. However, \method gradually and covertly infect the model with a backdoor through machine unlearning, i.e., the potential backdoor cannot be detected at the beginning.

\noindent \textbf{Data-Driven Machine Unlearning.} Machine unlearning is a field of research that deals with the process of removing or forgetting specific information stored in a machine learning model. The field has gained increasing attention due to the growing concern about privacy, security, and ethical implications associated with the use of machine learning models~\cite{machine_unlearning}. A straightforward approach to machine unlearning is to retrain the entire dataset, excluding the samples to be unlearned, however, it requires a lot of time and computational resources for large deep learning tasks. Therefore, Bourtoule et al.~\cite{sisa} introduced SISA framework using data partitioning techniques. It first divides the samples into shards, with each having an individual model, and the final outcome aggregates the predictions made by all the sub-models. Then each shard is further partitioned into slices, the intermediate model will be saved for each slice so that the model can be retrained from any state. In addition, data augmentation also can be used for machine unlearning. Huang et al.~\cite{error_minimizing} and Shan et al.~\cite{adv} respectively proposed to utilize the error-minimizing noise and targeted adversarial attack to prevent the model from learning from the targeted point. The rest group of unlearning algorithms first determines the impact of modifications in the training data on the parameters of the model, and subsequently unlearns the data by subtracting these effects. The impact can be evaluated through the influence functions~\cite{influence1,influence2} or Fisher Information Matrix~\cite{ssse}. Despite these advances, the field of machine unlearning is still in its early stages, and there are several open research challenges that need to be addressed, including the balance between efficiency and performance and the robustness and security of unlearning process.

\section{Conclusion}
\label{sec:conclusion}

In this work, we present a novel black-box backdoor attack algorithm, referred to as \methode, which is based on machine unlearning. \method involves implanting a backdoor into the model through a two-stage strategy. In the first phase, the adversary augments the clean training set with a set of carefully designed data, which consists of poison samples and mitigation samples, to obtain a `benign' initial model. In the second phase, the adversary continuously submits unlearning requests regarding the mitigation samples, thus gradually exposing the hidden backdoor in the initial model. Extensive experimental results demonstrate that in both exact unlearning and SISA scenarios, \method can successfully achieve the attack target, i.e., the initial model has a low attack success rate and is challenging to detect by advanced backdoor detection algorithms, however, after unlearning the mitigation samples, the model's attack success rate significantly increases. Moreover, we propose two sample-based defense methods for such attacks, and the experimental results indicate that they are effective in identifying whether the unlearning request is for a mitigation sample. Machine unlearning is a mechanism aimed at enhancing human trust in deep learning, especially its privacy. However, our work reveals that machine unlearning can pose new threats to deep learning.

\bibliographystyle{ACM-Reference-Format}
\bibliography{bdul}

\end{document}